\documentclass[twoside,preprintnumbers,amsmath,amssymb,pacs,nofootinbib]{revtex4}
\usepackage{amsmath,amssymb,marvosym}
\usepackage{graphicx}
\usepackage{braket,pslatex}

\renewcommand{\d}{\ensuremath{\mathrm{d}}}

\renewcommand{\d}{\ensuremath{\mathrm{d}}}

\newcommand{\p}{\partial}
\newcommand{\Evac}{E_{{\mbox{\tiny{vac}}}}}

\newcommand{\Q}{\ensuremath{\mathcal{Q}}}
\newcommand{\F}{\ensuremath{\mathcal{F}}}
\newcommand{\h}{\ensuremath{\mathcal{H}}}
\newcommand{\G}{\ensuremath{\mathcal{G}}}

\newcommand{\GZ}{\ensuremath{\mathrm{GZ}}}
\newcommand{\oG}{\ensuremath{\overline{\mathcal{G}}}}
\newcommand{\oF}{\ensuremath{\overline{\mathcal{F}}}}

\begin{document}

\title{{\Large { The Gribov horizon and spontaneous BRST symmetry breaking}}}

\author{David~Dudal}
\email{david.dudal@ugent.be}\ \thanks{Work supported by the Research-Foundation Flanders (FWO Vlaanderen).}
\affiliation{Ghent University, Department of Physics and Astronomy, Krijgslaan 281-S9, 9000 Gent, Belgium}
\author{Silvio Paolo Sorella}
\email{sorella@uerj.br}\ \thanks{Work supported by
FAPERJ, Funda{\c{c}}{\~{a}}o de Amparo {\`{a}} Pesquisa do Estado do Rio de
Janeiro, under the program \textit{Cientista do Nosso Estado}, E-26/101.578/2010.}
\affiliation{
UERJ - Universidade do Estado do Rio de Janeiro, Instituto de F\'{i}sica, Departamento de F\'{i}sica Te\'{o}rica, Rua S\~{a}o Francisco Xavier 524, 20550-013 Maracan\~{a}, Rio de Janeiro, RJ, Brasil}

\begin{abstract}
An equivalent formulation of the Gribov-Zwanziger theory accounting for the gauge fixing ambiguity in the Landau gauge is presented. The resulting action is constrained by a Slavnov-Taylor identity stemming from a nilpotent exact BRST invariance which is spontaneously broken due to the presence of the Gribov horizon. This spontaneous symmetry breaking can be described in a purely algebraic way through the introduction of a pair of auxiliary fields which give rise to a set of linearly broken Ward identities. The Goldstone sector turns out to be decoupled. The underlying exact nilpotent BRST invariance allows to employ BRST cohomology tools within the Gribov horizon to identify renormalizable extensions of gauge invariant operators. Using a simple toy model and appropriate Dirac bracket quantization, we discuss the time-evolution invariance of the operator cohomology.  We further comment on the unitarity issue in a confining theory, and stress that BRST cohomology alone is not sufficient to ensure unitarity, a fact, although well known, frequently ignored.
\end{abstract}

\maketitle
\section{Introduction}
The issue of the BRST symmetry in presence of the Gribov horizon is of pivotal relevance in order to achieve a better understanding of the infrared nonperturbative region of a confining and ultraviolet asymptotically free Yang-Mills theory. At the moment, it seems safe to state that reconciling BRST symmetry with the Gribov horizon is a task not yet fully  accomplished. As one learns from the original works of Gribov and Zwanziger \cite{Gribov:1977wm,Zwanziger:1989mf,Zwanziger:1992qr}, the restriction of the domain of integration in the functional Euclidean integral to the first Gribov horizon enables one to take into account the existence of the Gribov gauge fixing ambiguities in a local and renormalizable framework, as summarized by the Gribov-Zwanziger action in the Landau gauge \cite{Zwanziger:1989mf,Zwanziger:1992qr} and its more recent Refined version \cite{Dudal:2007cw,Dudal:2008sp,Dudal:2011gd}.

Two important physical consequences which arise from the restriction to the Gribov horizon are: the appearance of a nonperturbative dynamical mass parameter $\gamma^2\sim \Lambda_{\text{\tiny QCD}}^2$, called the Gribov mass, and a geometrical picture for the confinement of gluons.  Due to the presence of the Gribov horizon, the two-point gluon correlation function has no particle interpretation, displaying a violation of reflection positivity, a feature confirmed by numerical simulations \cite{Bowman:2007du}. It is worth mentioning that the expression for the tree level gluon propagator obtained from the Refined Gribov-Zwanziger (RGZ) theory \cite{Dudal:2007cw,Dudal:2008sp,Dudal:2011gd} is in very good agreement with the most recent lattice data on very large lattices \cite{Cucchieri:2007rg,Bornyakov:2008yx,Dudal:2010tf,Cucchieri:2011ig}. This propagator exhibits complex poles in momentum space\footnote{To avoid confusion, let us mention here that upon including perturbative corrections, the properties of the RGZ gluon propagator in the complex momentum plane will of course get affected.}, a feature which has a natural accommodation in terms of gluon confinement and absence of corresponding asymptotic states. Despite the appearance of complex poles, the aforementioned confining gluon propagator has been employed to investigate the correlation functions of gauge invariant composite glueball operators in order to get estimates of the glueball masses.  It turns out that a K{\"a}ll{\'e}n-Lehmann spectral representation with positive spectral density can be extracted from these correlation functions \cite{Baulieu:2009ha}. The resulting mass estimates for the lowest glueball states, $0^{++}, 2^{++}, 0^{-+},$ are in agreement with the available numerical data on the spectrum of the glueballs \cite{Dudal:2010cd}. It is worth mentioning that such type of gluon propagator has also been used in previous studies in hadron physics \cite{Roberts:1994dr}.

Although the above mentioned results can be interpreted as encouraging evidence of the fact that the Refined Gribov-Zwanziger theory can be effectively employed to investigate the physical spectrum of a confining Yang-Mills theory, we are far from having at our disposal a systematic way of constructing a set of physical composite operators whose correlation functions can be directly linked with the observable spectrum of the theory. In particular, the characterization of the spectral analytical properties and of the positivity of the corresponding correlation functions appears to be a highly cumbersome problem, taking into account that explicit calculations have to be done by employing a confining gluon propagator exhibiting complex poles. Let us point out here that also other continuum approaches can adequately reproduce the lattice results for the gluon propagator \cite{Cornwall:1981zr,Aguilar:2008xm,Fischer:2008uz,Tissier:2010ts,Boucaud:2011ug}, however the resulting  correlation functions  will also exhibit a complicated structure, so it is in any case unclear how these nontrivial propagators and eventually also vertex functions will conspire to form e.g.~a truly physical glueball correlation function.

We expect that a better understanding of the role of the BRST symmetry in confining gauge theories could shed some light on these issues. In the present paper we aim at presenting new results on the BRST symmetry in presence of the Gribov horizon\footnote{In the very recent paper \cite{Maas:2012ct}, it is mentioned that a Landau gauge has been constructed which involves averaging over all Gribov copies. The latter methodology would be an alternative to handling the Gribov copy problem compared to the restriction within the Gribov horizon. Again according to \cite{Maas:2012ct}, it has been shown in some of the quoted literature that such averaging harbors an intact BRST symmetry. We have however been unable to recover from that quoted literature either how to average over all Gribov copies in the continuum, or why this would induce a well-defined BRST symmetry. In fact, a recent continuum attempt towards sampling over all Gribov copies in \cite{Serreau:2012cg} seems to give rise to a Lagrangian without BRST invariance, given the emerging connection with the Curci-Ferrari model, which is known to not exhibit exact BRST invariance. A similar observation was made in \cite{vonSmekal:2008en}: ``Only if the numerical lattice procedure converges towards a correct sampling of the fundamental
modular region in the infinite volume and continuum limits, can the BRST breaking effects be expected to go away.''. Clearly, the disappearance of BRST breaking effects in the continuum is to date thus merely a wishful expectation, being an extremely hard problem to actually solve. No continuum approach to restricting the functional integration to this fundamental modular region \cite{vanBaal:1991zw} is known. }. For the benefit of the reader, let us proceed by providing a short summary of previous attempts on this topic. We may start by remarking that already in the original paper by Zwanziger on the renormalizability of the Gribov-Zwanziger theory \cite{Zwanziger:1989mf,Zwanziger:1992qr}, it was pointed out that the restriction to the Gribov horizon induces a soft breaking of the BRST symmetry which does not jeopardize the renormalizability of the Gribov-Zwanziger action. Later on, in \cite{Maggiore:1993wq}, an attempt was made to argue that the Gribov-Zwanziger action displays an exact BRST invariance which is, however, spontaneously broken. Moreover, the spontaneous symmetry breaking was encoded into the BRST transformations through the introduction of terms containing an explicit dependence from the space-time coordinates $x_\mu$, see \cite{Vandersickel:2012tz} for more details. As pointed out in \cite{Dudal:2008sp}, this kind of $x_\mu$-dependent transformations is plagued by several drawbacks as, for example: the explicit loss of  translational invariance and the appearance of nontrivial surface terms requiring appropriate boundary conditions for the fields at infinity. In \cite{Baulieu:2008fy}, the soft breaking induced by the Gribov horizon was discussed in terms of the BRST cohomology. It was shown that the soft breaking enables to give a cohomological proof of the fact that the Gribov mass $\gamma^2$ is a physical relevant parameter. Also, the idea of the soft BRST breaking was employed to construct a renormalizable quark model \cite{Baulieu:2009xr} accounting for the lattice data on the quark propagator \cite{Parappilly:2005ei,Furui:2006ks}. Further, in  \cite{Sorella:2009vt} it was pointed out that the soft breaking induced by the Gribov horizon could be converted into an exact invariance which, however, gives rise to a non-local and non-nilpotent BRST operator.  This non-local formulation of the BRST invariance  was subsequently casted in a local framework in  \cite{Dudal:2010hj}. Though, the resulting BRST operator still lacks nilpotency. We mention that in  \cite{Dudal:2009xh} a connection between the soft BRST breaking appearing in the Gribov-Zwanziger theory and the Kugo-Ojima criterion \cite{Kugo:1979gm} was established. More recently, the Gribov-Zwanziger theory was reformulated in such a way that the soft BRST breaking was transformed into a linear breaking \cite{Capri:2010hb,Capri:2011wp}, a feature which allowed to clarify the nonrenormalizability properties of the Gribov parameter $\gamma^2$.

In this work we shall present a new equivalent formulation of the Gribov-Zwanziger action which exhibits an exact BRST invariance, with a nilpotent BRST operator. Moreover, the BRST symmetry is spontaneously broken, the spontaneous breaking parameter being nothing but the Gribov mass $\gamma^2$. However, unlike the previous formulation of  \cite{Maggiore:1993wq}, no terms with explicit $x_\mu$-dependence are introduced in the BRST transformations of the fields. This is an important feature of this new formulation, allowing for a fully local, covariant and renormalizable framework. Another important outcome is the decoupling of the Goldstone mode associated to the spontaneous symmetry breaking. Moreover, the nilpotency of the BRST symmetry enables us to make use of cohomology tools to identify a set of local and renormalizable operators in presence of the Gribov horizon. The  paper is organized as follows. Section II is devoted to the presentation of the new equivalent formulation of the Gribov-Zwanziger action and of the spontaneous symmetry breaking of the BRST operator.  In Section III we collect the set of Ward identities fulfilled by the reformulated action and we analyze the decoupling of the Goldstone mode. In Section IV we discuss,  with the help of a toy model, the invariance under time evolution of the set of composite operators corresponding to the cohomology classes of the BRST operator. Section V collects our conclusion.

\section{Reformulation of the Gribov-Zwanziger action and spontaneous BRST symmetry breaking}
\subsection{An equivalent reformulation of the Gribov-Zwanziger action}
Our starting point is the Gribov-Zwanziger action which enables us to restrict the Euclidean functional integral to the first Gribov horizon, namely
\begin{eqnarray}\label{GZaction}
S_\GZ &=& \frac{1}{4}\int \d^d x\; F^a_{\mu\nu} F^a_{\mu\nu}+\int \d^d x\,\left( b^a \p_\mu A_\mu^a +\overline c^a \p_\mu D_\mu^{ab} c^b \right)+\int \d^d x \left( \overline{\varphi }_{\mu }^{ac} \partial _{\nu} D_\nu^{am}\varphi _{\mu
}^{mc} -\overline{\omega }_{\mu }^{ac} \partial _{\nu } D_\nu^{am} \omega _{\mu }^{mc}
  -g\left( \partial _{\nu }\overline{\omega }_{\mu}^{ac}\right) f^{abm}\left( D_{\nu }c\right) ^{b}\varphi _{\mu
}^{mc}\right)\nonumber\\
&&+\int\d^d x\left( -\gamma ^{2}gf^{abc}A_{\mu }^{a}(\varphi _{\mu }^{bc}+\overline{\varphi }_{\mu }^{bc}) - d\left(N^{2}-1\right) \gamma^{4} \right) \;.
\end{eqnarray}
The Gribov parameter  $\gamma^2$ is not a free mass, but it is determined in a self-consistent way through the gap equation $\left.\frac{\p \Evac}{\p \gamma^2}\right|_{\gamma^2\neq0}=0$, where $\Evac$ stands for the vacuum energy of the theory \cite{Gribov:1977wm,Zwanziger:1989mf,Zwanziger:1992qr}.

The foregoing model has a softly broken BRST invariance, generated by the nilpotent transformations
\begin{eqnarray}
sA_{\mu }^{a}& =&-D_{\mu }^{ab}c^{b}=-(\partial _{\mu }\delta
^{ab}+gf^{acb}A_{\mu }^{c})c^{b}\;,\qquad sc^{a}~=~\frac{g}{2}f^{acb}c^{b}c^{c}\;,\qquad s\overline{c}^{a}~=~b^{a}\;,\qquad sb^{a}~=~0\;, \nonumber \\
s\overline{\omega}_{\mu }^{ab}& =&\overline{\varphi}_{\mu }^{ab}\;,\qquad s\overline{\varphi}_{\mu }^{ab}~=~0\;,\qquad s\varphi _{\mu }^{ab}~=~\omega _{\mu }^{ab}\;,\qquad s\omega _{\mu
}^{ab}~=~0\;.  \label{BRS}
\end{eqnarray}
Indeed, we have $s^2=0$ while
\begin{eqnarray}
sS_\GZ&=& -\gamma^2\int \d^d x\left(gf^{abc}D_\mu^{ak}c^k(\overline\varphi_\mu^{bc}+\varphi_\mu^{bc})+gf^{abc}A_\mu^a\omega_\mu^{bc}\right)  \;.
\end{eqnarray}
We propose to replace expression \eqref{GZaction} by the following action
\begin{eqnarray}\label{GZaction2}
S'_\GZ &=& \frac{1}{4}\int \d^d x\; F^a_{\mu\nu} F^a_{\mu\nu}+\int \d^d x\,\left( b^a \p_\mu A_\mu^a +\overline c^a \p_\mu D_\mu^{ab} c^b \right)\nonumber\\
&&+\int \d^d x \left( \overline{\varphi }_{\mu }^{ac} \partial _{\nu} D_\nu^{am}\varphi _{\mu
}^{mc} -\overline{\omega }_{\mu }^{ac} \partial _{\nu } D_\nu^{am} \omega _{\mu }^{mc}
  -g\left( \partial _{\nu }\overline{\omega }_{\mu}^{ac}\right) f^{abm}\left( D_{\nu }c\right) ^{b}\varphi _{\mu
}^{mc}\right)\nonumber\\
&&+\int\d^d x\left(- \oG_{\mu\nu}^{ab}\p^2\G_{\mu\nu}^{ab}+ \oF_{\mu\nu}^{ab}\p^2\F_{\mu\nu}^{ab}- \oG_{\mu\nu}^{ab}D_{\mu}^{ak}\overline\varphi_\nu^{kb}-gf^{ak\ell }\oG_{\mu\nu}^{ab}D_\mu^{\ell p}c^p\overline\omega_\nu^{k b}  \right) \nonumber\\
&&+\int\d^d x\left(- \widehat{\oG}_{\mu\nu}^{ab}\p^2\widehat{\G}_{\mu\nu}^{ab}+ \widehat{\oF}_{\mu\nu}^{ab}\p^2\widehat{\F}_{\mu\nu}^{ab}+\widehat{\oF}_{\mu\nu}^{ab}D_\mu^{ak}\omega^{kb}_\nu- \widehat{\oG}_{\mu\nu}^{ab}D_{\mu}^{ak}\varphi_\nu^{kb}+gf^{a k\ell}\widehat{\oF}_{\mu\nu}^{ab}D_\mu^{\ell p}c^p\varphi_\nu^{k b} \right)\nonumber\\&&
+\int\d^d x\left(\h_{\mu\nu}^{ab}\left(\oG_{\mu\nu}^{ab}-\delta_{\mu\nu}\delta^{ab}\gamma^2\right)+\widehat{\h}_{\mu\nu}^{ab}\left(\widehat{\oG}_{\mu\nu}^{ab}-\delta_{\mu\nu}\delta^{ab}\gamma^2\right)-\oG_{\mu\nu}^{ab}\widehat{\oG}_{\mu\nu}^{ab}\right)
\end{eqnarray}
by introducing two new BRST quartets of fields\footnote{We dressed these new fields with the designated color and Lorentz-structure as this formulation allows for several Ward identities, which will be employed to prove the all orders renormalizability of the action  $S'_\GZ$ in a follow-up paper.} consisting of $\G_{\mu\nu}^{ab}$, $\oG_{\mu\nu}^{ab}$ (commuting),  $\F_{\mu\nu}^{ab}$, $\oF_{\mu\nu}^{ab}$ (anticommuting) and their hat-counterparts:
\begin{eqnarray}\label{quartet1}
    s\overline \F_{\mu\nu}^{ab}&=&\oG_{\mu\nu}^{ab}\,,\qquad s \oG_{\mu\nu}^{ab} ~=~ 0\,,\qquad s \widehat{\oF}_{\mu\nu}^{ab}=\widehat{\oG}_{\mu\nu}^{ab}\,,\qquad s \widehat{\oG}_{\mu\nu}^{ab} ~=~ 0\,,\nonumber\\
    s \G_{\mu\nu}^{ab}&=&\F_{\mu\nu}^{ab}\,,\qquad s \F_{\mu\nu}^{ab}~=~0\,,\qquad  s\widehat{\G}_{\mu\nu}^{ab}=\widehat{\F}_{\mu\nu}^{ab}\,,\qquad s \widehat{\F}_{\mu\nu}^{ab}~=~0  \label{BRS1}
\end{eqnarray}
and the singlets $\h_{\mu\nu}^{ab},\widehat{\h}_{\mu\nu}^{ab}$, with thus $s\h_{\mu\nu}^{ab}=s\widehat{\h}_{\mu\nu}^{ab}=0$. As the action \eqref{GZaction2} can be easily rewritten as
\begin{eqnarray}\label{GZaction2bis}
S'_\GZ &=& \frac{1}{4}\int \d^d x\; F^a_{\mu\nu} F^a_{\mu\nu}+\int \d^d x\,s\left( \overline c^a \p_\mu A_\mu^a +\overline{\omega }_{\mu }^{ac} \partial _{\nu} D_\nu^{am}\varphi _{\mu
}^{mc}\right)\nonumber\\
&&+\int\d^d x\,s\left(- \oF_{\mu\nu}^{ab}\p^2\G_{\mu\nu}^{ab}- \oG_{\mu\nu}^{ab}D_{\mu}^{ak}\overline\omega_\nu^{kb}  - \widehat{\oF}_{\mu\nu}^{ab}\p^2\widehat{\G}_{\mu\nu}^{ab}- \widehat{\oF}_{\mu\nu}^{ab}D_{\mu}^{ak}\varphi_\nu^{kb} \right)\nonumber\\&&
+\int\d^d x\left(\h_{\mu\nu}^{ab}\left(\oG_{\mu\nu}^{ab}-\delta_{\mu\nu}\delta^{ab}\gamma^2\right)+\widehat{\h}_{\mu\nu}^{ab}\left(\widehat{\oG}_{\mu\nu}^{ab}-\delta_{\mu\nu}\delta^{ab}\gamma^2\right)-\oG_{\mu\nu}^{ab}\widehat{\oG}_{\mu\nu}^{ab}\right)\;,
\end{eqnarray}
the BRST invariance is almost immediate, $sS'_{GZ}=0$, whereby we have preserved the nilpotency of the BRST operator, $s^2=0$. The role of the fields $(\h,\widehat{\h})$ is clearly seen, as upon using their algebraic exact equations of motion (EOM), $\oG_{\mu\nu}^{ab}=\widehat{\oG}_{\mu\nu}^{ab}=\gamma^2\delta^{ab}\delta_{\mu\nu}$, we get as an immediate consequence
\begin{eqnarray}\label{GZaction2tris}
&&\int\d^d x\left(- \oG_{\mu\nu}^{ab}D_{\mu}^{ak}\overline\varphi_\nu^{kb}- \widehat{\oG}_{\mu\nu}^{ab}D_{\mu}^{ak}\varphi_\nu^{kb}- \oG_{\mu\nu}^{ab}\p^2\G_{\mu\nu}^{ab}- \widehat{\oG}_{\mu\nu}^{ab}\p^2\widehat{\G}_{\mu\nu}^{ab}-\oG_{\mu\nu}^{ab}\widehat{\oG}_{\mu\nu}^{ab}+\h_{\mu\nu}^{ab}\left(\oG_{\mu\nu}^{ab}-\delta_{\mu\nu}\delta^{ab}\gamma^2\right)\right)\nonumber\\
&&\stackrel{\h-,\widehat{\h}-EOM}{\longrightarrow}
\int\d^d x\left(- \gamma^2 gf^{abc}A_\mu^a \left(\varphi_\mu^{bc}+\overline\varphi_\mu^{bc}\right)-d(N^2-1)\gamma^4\right)\,.
\end{eqnarray}
Naively, the doublets $(\oF,\oG)$, $(\widehat{\oF},\widehat{\oG})$  together with the $(\h,\widehat{\h})$-terms would be sufficient to write down a BRST-invariant action that reduces, upon using the $(\h,\widehat{\h})$-EOM, to the original GZ action \eqref{GZaction}. However, this would leave us with a trivial partition function, due to the vanishing path integration over $[\d \F]$ or $[\d \widehat{\F}]$. The use of the double quartet circumvents this problem, neither corrupting the BRST invariance nor the exact correspondence with the action \eqref{GZaction}. In fact, the integration over $\F$, $\oF$, $\widehat{\F}$ and $\widehat{\oF}$ will now generate a unity, due to absence of any other term in $\F$ or $\widehat{\F}$, thereby removing all terms $\sim\oF$ and $\sim \widehat{\oF}$. This can be best seen by performing 3 simultaneous shifts of integration variables:
\begin{eqnarray}
\widehat{\F}_{\mu\nu}^{ab}&\to&\widehat{\F}_{\mu\nu}^{ab}-\frac{1}{\p^2}\left(D_\mu^{ak}\omega_{\nu}^{kb}+gf^{a k\ell}D_\mu^{\ell p}c^p\varphi_\nu^{k b}\right)\,,\qquad
\omega_{\nu}^{tb}~\to~\omega_{\nu}^{tb}+\left[(\p D)^{-1}\right]^{tk}\left(gf^{ak\ell} \oG_{\mu\nu}^{ab}D_\mu^{\ell p}c^p\right)\,,\nonumber\\
\F_{\mu\nu}^{ab}&\to&\F_{\mu\nu}^{ab}-\frac{1}{\p^2}\left(D_\mu^{ak}\omega_\nu^{kb}+gf^{a k\ell}D_\mu^{\ell p}c^p\overline\varphi_\nu^{k b}\right) \;,
\end{eqnarray}
from which the equivalence between the two formulations, \eqref{GZaction} and \eqref{GZaction2}, becomes apparent.

\subsection{Spontaneous symmetry breaking of the BRST symmetry}
In order to realize that the new formulation of the Gribov-Zwanziger action does exhibit a spontaneous symmetry breaking of the BRST symmetry, we rewrite the action $S'_\GZ$ by making explicit use of the  equations of motion of the auxiliary fields $(\h,\widehat{\h})$:
\begin{equation}
\oG_{\mu\nu}^{ab}=\widehat{\oG}_{\mu\nu}^{ab}=\gamma^2\delta^{ab}\delta_{\mu\nu} \;. \label{heom}
\end{equation}
Thus
\begin{eqnarray}\label{GZactioneom}
S'_\GZ &=& \frac{1}{4}\int \d^d x\; F^a_{\mu\nu} F^a_{\mu\nu}+\int \d^d x\,s\left( \overline c^a \p_\mu A_\mu^a +\overline{\omega }_{\mu }^{ac} \partial _{\nu} D_\nu^{am}\varphi _{\mu
}^{mc}\right)\nonumber\\
&&+\int\d^d x\;  \left( \oF_{\mu\nu}^{ab}\p^2\F_{\mu\nu}^{ab}-   \gamma^2 \; s (D_{\mu}^{ak}\overline\omega_\mu^{ka})
 + \widehat{\oF}_{\mu\nu}^{ab}\p^2\widehat{\F}_{\mu\nu}^{ab}- \gamma^2 D_{\mu}^{ak}\varphi_\mu^{ka}
 +  \widehat{\oF}_{\mu\nu}^{ab} \;s (D_{\mu}^{ak}\varphi_\nu^{kb})  - \gamma^4 d (N^2-1)  \right) \;.
\end{eqnarray}
Expression \eqref{GZactioneom} is easily seen to be left invariant by the following nilpotent BRST transformations:
\begin{eqnarray}
sA_{\mu }^{a}& =&-D_{\mu }^{ab}c^{b}=-(\partial _{\mu }\delta
^{ab}+gf^{acb}A_{\mu }^{c})c^{b}\;,  \qquad sc^{a}~=~\frac{g}{2}f^{acb}c^{b}c^{c}\;,\qquad s\overline{c}^{a}~=~b^{a}\;,\qquad sb^{a}~=~0\;,  \nonumber \\
s\overline{\omega}_{\mu }^{ab}& =&\overline{\varphi}_{\mu }^{ab}\;,\qquad s\overline{\varphi}%
_{\mu }^{ab}~=~0\;,  \qquad s\varphi _{\mu }^{ab}~=~\omega _{\mu }^{ab}\;,\qquad s\omega _{\mu
}^{ab}~=~0\;,   \nonumber\\
s\oF_{\mu\nu}^{ab} & =& \gamma^2 \delta^{ab}\delta_{\mu\nu}\;, \qquad  s  \widehat{\oF}_{\mu\nu}^{ab}~=~ \gamma^2 \delta^{ab}\delta_{\mu\nu} \;, \qquad s \G_{\mu\nu}^{ab}~=~ \F_{\mu\nu}^{ab}\;, \qquad s \F_{\mu\nu}^{ab}~=~ 0\;, \qquad s\widehat{\G}_{\mu\nu}^{ab}~=~\widehat{\F}_{\mu\nu}^{ab}\,,\qquad s \widehat{\F}_{\mu\nu}^{ab}~=~0
\label{BRSTssb}
\end{eqnarray}
with
\begin{equation}
s S'_\GZ = 0\,,\qquad s^2  =0 \;. \label{ssbinv}
\end{equation}
Moreover, from equations \eqref{BRSTssb}, it follows that the BRST operator suffers from spontaneous symmetry breaking, namely
\begin{equation}
\langle s\oF_{\mu\nu}^{ab} \rangle  = \gamma^2 \delta^{ab}\delta_{\mu\nu} \;, \qquad  \langle s  \widehat{\oF}_{\mu\nu}^{ab} \rangle = \gamma^2 \delta^{ab}\delta_{\mu\nu} \;.  \label{s_ssb}
\end{equation}
A few remarks are in order:
\begin{itemize}
\item unlike the formulation of \cite{Maggiore:1993wq}, terms with explicit dependence from the space-time coordinates $x_\mu$ are introduced neither in the BRST transformations \eqref{BRSTssb} nor in eqs.\eqref{s_ssb}. This relevant feature enables us to study the spontaneous breaking within the well-established framework of genuine local quantum field theory. Indeed, one notices that both the action $S'_\GZ$, eq.\eqref{GZaction2}, and the BRST transformations, eqs.\eqref{BRSTssb}, are local.

\item as it is apparent from eqs.\eqref{s_ssb},  the spontaneous breaking of the BRST symmetry is entirely driven by the Gribov mass parameter $\gamma^2$. This means that the breaking is due to the restriction to the Gribov horizon. As such, it is a truly non-perturbative effect with a clear geometrical meaning.

\item it is worth to point out that the introduction of the auxiliary fields $(\h,\widehat{\h})$ enables us to treat the spontaneous symmetry breaking in a purely algebraic way, as it follows from the equations of motion of these fields:
\begin{equation}
\frac{\delta S'_\GZ}{\delta \h_{\mu\nu}^{ab}} = \oG_{\mu\nu}^{ab}-\delta_{\mu\nu}\delta^{ab}\gamma^2  \;, \qquad \frac{\delta S'_\GZ}{\delta \widehat{\h}_{\mu\nu}^{ab}} = \widehat{\oG}_{\mu\nu}^{ab}-\delta_{\mu\nu}\delta^{ab}\gamma^2 \;.  \label{linh}
\end{equation}
In fact, according to the algebraic renormalization procedure \cite{Piguet:1995er}, equations \eqref{linh} will acquire the meaning of linearly broken Ward identities at quantum level, providing a very powerful and elegant treatment of the properties of the spontaneous symmetry breaking at the quantum level, see next Section,
\item although not worked out in explicit way here, we underline that the spontaneous symmetry breaking mechanism presented in this work  generalizes without any difficulty to the Refined Gribov-Zwanziger action \cite{Dudal:2007cw,Dudal:2008sp,Dudal:2011gd}. Also a soft breaking in the fermion sector \cite{Baulieu:2008fy,Baulieu:2009xr} fits within this new scheme.
\end{itemize}

\section{Restrictions on the effective action}
\subsection{Set of Ward identities}
In a follow-up paper \cite{prep}, we will discuss the renormalizability of the new reformulated  GZ action \eqref{GZaction2}. For that purpose, let us already list here a set  of Ward identities that will be helpful to construct the most general admissible counterterm. According to the algebraic renormalization setup  \cite{Piguet:1995er}, we have to couple the nonlinear BRST variation of $A_\mu^a$ and $c^a$ to BRST invariant sources $K_\mu^a$ and $L^a$. In addition, following \cite{Zwanziger:1989mf,Zwanziger:1992qr}, we can introduce a composite index notation, $i=(\nu,b)$, as follows
\begin{eqnarray}
  \chi_{\mu\nu}^{ab} &=& \chi_{\mu i}^a\,,\qquad   \eta_{\nu}^{ab} ~=~ \eta_{i}^a \;,
\end{eqnarray}
where $ \chi_{\mu\nu}^{ab},  \eta_{\nu}^{ab}$ stand for generic fields of the action \eqref{GZaction2}.
Also, for the identification of two more powerful Ward identities, we have found it useful to introduce a further BRST doublet of external sources
\begin{equation}\label{extradoub}
    s\lambda_i^{ab}~=~\rho_i^{ab}\,,\qquad s\rho_i^{ab}~=~0
\end{equation}
coupled to the composite operator $\widehat\oF_{\mu i}^a D_\mu^{bk}c^k$ and its BRST variation $\widehat\oG_{\mu i}^aD_\mu^{bk}c^k$.

We thus look at  Ward identities for the complete classical BRST invariant action
 \begin{eqnarray}\label{GZaction2bis}
\Sigma &=&\frac{1}{4}\int \d^d x\; F^a_{\mu\nu} F^a_{\mu\nu}+\int \d^d x\,\left( b^a \p_\mu A_\mu^a +\overline c^a \p_\mu D_\mu^{ab} c^b \right)\nonumber\\
&&+\int \d^d x \left( \overline{\varphi }_{i }^{a} \partial _{\nu} D_\nu^{am}\varphi _{i
}^{m} -\overline{\omega }_{i}^{a} \partial _{\nu } D_\nu^{am} \omega _{i }^{m}
  -g\left( \partial _{\nu }\overline{\omega }_{i}^{a}\right) f^{abm} D_{\nu }^{bk}c^{k}\varphi _{i
}^{m}\right)\nonumber\\
&&+\int\d^d x\left(- \oG_{\mu i}^{a}\p^2\G_{\mu i}^{a}+ \oF_{\mu i}^{a}\p^2\F_{\mu i}^{a}- \oG_{\mu i}^{a}D_{\mu}^{ak}\overline\varphi_i^{k}-gf^{a k\ell}\oG_{\mu i}^{a}D_\mu^{\ell p}c^p\overline\omega_i^{k}  \right) \nonumber\\
&&+\int\d^d x\left(- \widehat{\oG}_{\mu i}^{a}\p^2\widehat{\G}_{\mu i}^{a}+ \widehat{\oF}_{\mu i}^{a}\p^2\widehat{\F}_{\mu i}^{a}+\widehat{\oF}_{\mu i}^{a}D_\mu^{ak}\omega^{k}_i- \widehat{\oG}_{\mu i}^{a}D_{\mu}^{ak}\varphi_i^{k}+gf^{ak\ell}\widehat{\oF}_{\mu i}^{a}D_\mu^{\ell p}c^p\varphi_i^{k } \right)\nonumber\\&&
+\int\d^d x\left(\h_{\mu i}^{a}\left(\oG_{\mu i}^{a}-\delta_{\mu i}^{a}\gamma^2\right)+\widehat{\h}_{\mu i}^{a}\left(\widehat{\oG}_{\mu i}^{a}-\delta_{\mu i}^{a}\gamma^2\right)-\oG_{\mu i}^{a}\widehat{\oG}_{\mu i}^{a}\right)\nonumber\\&& + \int\d^d x \left( -K_\mu^a D_\mu^{ab}c^b + \frac{g}{2}L^a f^{abc}c^b c^c+\rho_i^{ab}\widehat\oG_{\mu i}^aD_\mu^{bk}c^k-\lambda_i^{ab}\widehat\oF_{\mu i}^aD_\mu^{bk}c^k\right)\;.
\end{eqnarray}
It turns out that this action  $\Sigma$ fulfills the following Ward identities:
\begin{enumerate}
\item \emph{An exact Slavnov-Taylor identity}
\begin{eqnarray}
  \int \d^dx \left(\frac{\delta \Sigma}{\delta K_\mu^a}\frac{\delta \Sigma}{\delta A_\mu^a}+\frac{\delta \Sigma}{\delta L^a}\frac{\delta \Sigma}{\delta c^a}+b^a\frac{\delta \Sigma}{\delta \overline c^a}+\overline\varphi_i^a\frac{\delta \Sigma}{\delta \overline\omega_i^a}+\omega_i^a\frac{\delta \Sigma}{\delta \varphi_i^a}+\oG_{\mu i}^a\frac{\delta \Sigma}{\delta \oF_{\mu i}^a}+\F_{\mu i}^a\frac{\delta \Sigma}{\delta \G_{\mu i}^a}+\widehat{\oG}_{\mu i}^a\frac{\delta \Sigma}{\delta \widehat{\oF}_{\mu i}^a}+\widehat{\F}_{\mu i}^a\frac{\delta \Sigma}{\delta \widehat{\G}_{\mu i}^a} +\rho_i^{ab}\frac{\delta}{\delta \lambda_i^{ab}}\right) &=& 0\,.\nonumber\\
\end{eqnarray}
\item \emph{The linearly broken Ward identity for the Gribov mass parameter $\gamma^2$}
\begin{eqnarray}
\frac{\delta \Sigma}{\delta \gamma^2}&=& -\int d^4x\left(\h_{\mu\mu}^{aa}+\widehat{\h}_{\mu\mu}^{aa}\right)\,.
\end{eqnarray}
\item \emph{The linearly broken equations-of-motion identities}
\begin{eqnarray}
\frac{\delta \Sigma}{\delta b^a}&=& \p_\mu A_\mu^a\,,\qquad \frac{\delta \Sigma}{\delta \overline c^a}+\p_\mu\frac{\delta \Sigma}{\delta K_\mu^a}~=~0\,,\\
\frac{\delta \Sigma}{\delta \G_{\mu i}^a}&=& -\p^2 \oG_{\mu i}^{a}\,,\qquad \frac{\delta \Sigma}{\delta \widehat{\G}_{\mu i^a}}~=~ -\p^2 \widehat{\oG}_{\mu i}^{a}\,,\label{ward1}\\
\frac{\delta \Sigma}{\delta \F_{\mu i}^a}&=& -\p^2 \oF_{\mu i}^{a}\,,\qquad \frac{\delta \Sigma}{\delta \oF_{\mu i}^a}~=~ \p^2 \F_{\mu i}^{a}\,,\qquad\frac{\delta \Sigma}{\delta \widehat{\F}_{\mu i^a}}~=~ -\p^2 \widehat{\oF}_{\mu i}^{a}\,,\label{nodig}\\
\frac{\delta \Sigma}{\delta \h_{\mu i}^a}&=& \oG_{\mu i}^a-\gamma^2\delta_{\mu i}^a\,,\qquad \frac{\delta \Sigma}{\delta \widehat{\h}_{\mu i}^a}~=~ \widehat{\oG}_{\mu i}^a-\gamma^2\delta_{\mu i}^a\,.\label{ward2}
\end{eqnarray}
\item \emph{Other linearly broken equations-of-motion type identities}
\begin{eqnarray}
  \frac{\delta \Sigma}{\delta \overline\varphi_i^a}+\p_\nu\frac{\delta \Sigma}{\delta \oG_{\nu i}^a}-D_\nu^{ak}\frac{\delta \Sigma}{\delta \h_{\nu i}^k} &=& -\p^2 \p_\nu\widehat\G_{\nu i}^a+\p_\nu \widehat\h_{\nu i}^a-\p_\nu \oG_{\nu i}^a  -gf^{ak\ell}A_\nu^\ell \delta_{\nu i}^k\gamma^2\,,\\
  \frac{\delta \Sigma}{\delta \overline\omega_i^a}+\p_\mu\frac{\delta \Sigma}{\delta \widehat\oF_{\mu i}^a}-g f^{ak\ell}\left(\frac{\delta \Sigma}{\delta \h_{\mu i}^k}+\delta_{\mu i}^k\gamma^2\right)\frac{\delta \Sigma}{\delta K_\mu^\ell}+\p_\mu\left(\lambda_{i}^{ab}\frac{\delta \Sigma}{\delta K_\mu^b}\right)&=&\p^2\p_\mu \widehat\F_{\mu i}^a\,.
\end{eqnarray}
\item \emph{The linearly broken integrated $\varphi$-equation-of-motion}
\begin{eqnarray}
  \int d^4x\left(\frac{\delta \Sigma}{\delta \varphi_i^a}+gf^{ka\ell}\overline\varphi_i^k\frac{\delta \Sigma}{\delta b^\ell}-gf^{mba}\overline\omega_i^m\frac{\delta \Sigma}{\delta \overline c^b}-gf^{ka\ell}A_\mu^\ell \frac{\delta \Sigma}{\delta \h_{\mu i}^k}+gf^{ka\ell}\frac{\delta \Sigma}{\delta \lambda_{i}^{k\ell}}\right)&=&\gamma^2\delta_{\mu i}^kgf^{ka\ell}\int d^4x  A_\mu^\ell\,.
\end{eqnarray}
\item \emph{The exact integrated Ward identities, generated by:}
\begin{eqnarray}
  \left(\oF_{\mu i}^a \frac{\delta}{\delta \G_{\mu j}^a}-\oG_{\mu j}^a \frac{\delta}{\delta \F_{\mu i}^a}\right)\Sigma &=& 0\,,\qquad
  \left(\widehat{\oF}_{\mu i}^a \frac{\delta}{\delta \widehat{\G}_{\mu j}^a}-\widehat{\oG}_{\mu j}^a \frac{\delta}{\delta \widehat{\F}_{\mu i}^a}\right)\Sigma ~=~ 0\,,\nonumber\\
  \left(\widehat{\oF}_{\mu i}^a \frac{\delta}{\delta \G_{\mu j}^a}-\oG_{\mu j}^a \frac{\delta}{\delta \widehat{\F}_{\mu i}^a}\right)\Sigma &=& 0\,,\qquad
  \left(\oF_{\mu i}^a \frac{\delta}{\delta \widehat{\G}_{\mu j}^a}-\widehat{\oG}_{\mu j}^a \frac{\delta}{\delta \F_{\mu i}^a}\right)\Sigma ~=~ 0\,.
\end{eqnarray}
\item \emph{ The linearly broken integrated global $Q_{ij}$ Ward identity}
  \begin{eqnarray}
Q_{ij}\Sigma&\equiv&  \int d^4x \left(\varphi_i^a\frac{\delta}{\delta \varphi_j^a}-\overline\varphi_j^a\frac{\delta}{\delta \overline \varphi_i^a}+\omega_i^a\frac{\delta}{\delta \omega_j^a}-\overline\omega_j^a\frac{\delta}{\delta \overline \omega_i^a}+\oG_{\mu i}^a\frac{\delta}{\delta \oG_{\mu j}^a}-\G_{\mu j}^a\frac{\delta}{\delta \G_{\mu i}^a}+\oF_i^a\frac{\delta}{\delta \oF_j^a}-\F_j^a\frac{\delta}{\delta \F_i^a}\right.\nonumber\\&&\left.-\widehat\oG_i^a\frac{\delta}{\delta \widehat\oG_j^a}+\widehat\G_j^a\frac{\delta}{\delta \widehat\G_i^a}-\widehat\oF_i^a\frac{\delta}{\delta \widehat\oF_j^a}+\widehat\F_j^a\frac{\delta}{\delta \widehat\F_i^a}-\h_{\mu j}^a\frac{\delta}{\delta \h_{\mu i}^a}+\widehat\h_{\mu j}^a\frac{\delta}{\delta \widehat\h_{\mu i}^a}+\rho_j^{ab}\frac{\delta }{\delta \rho_i^{ab}}+\lambda_j^{ab}\frac{\delta }{\delta \lambda_i^{ab}}\right)\Sigma=0\,.
\end{eqnarray}
\item \emph{An integrated Ward identity involving the Faddeev-Popov ghosts and the Gribov-Zwanziger fields}
 \begin{eqnarray}
    \int d^4x \left(c^a\frac{\delta \Sigma}{\delta \omega_i^a}-\overline\omega_i^a\frac{\delta \Sigma}{\delta \overline c^a}+\delta^{ab}\frac{\delta \Sigma}{\delta \lambda_{\mu}^{ab}}\right)&=&0\,.
    \end{eqnarray}
\end{enumerate}
All these identities are compatible with the Quantum Action Principle \cite{Piguet:1995er}. As such, they will severely constrain the possible counterterm which may arise at the quantum level. Although the output  of the algebraic renormalization analysis will be presented in a more complete and detailed work, we expect that this set of Ward identities will ensure the multiplicative renormalizability of the reformulated GZ action \eqref{GZaction2}, as  supported by the physical equivalence between the new and the original formulations of the GZ theory. In particular, this will imply that the quantum effective action, $\Gamma = \Sigma + O(\hbar)$, will fulfill the same set of  Ward identities obeyed by $\Sigma$.
\subsection{Decoupling of BRST Goldstone mode}
The exact quantum EOMs, $0=\frac{\delta \Gamma}{\delta \h_{\mu i}^a}=\oG_{\mu i}^a-\gamma^2\delta_{\mu i}^a$, $0=\frac{\delta \Gamma}{\delta \widehat\h_{\mu i}^a}=\widehat\oG_{\mu i}^a-\gamma^2\delta_{\mu i}^a$ will lead to $\oG_{\mu i}^a=\widehat\oG_{\mu i}^a=\gamma^2\delta_{\mu i}^a$, and thus to the spontaneous breakdown of the continuous BRST symmetry, henceforth a Goldstone degree of freedom must emerge. The latter can be created out from  the vacuum at no energy cost using the BRST current. Writing down here only the relevant part of this current
\begin{equation}\label{currentbrst}
    \mathcal{J}_\mu^{BRST}=\ldots+\oG_{\nu i}^a \p_\mu \F_{\nu i}^a+\widehat\oG_{\nu i}^a \p_\mu \widehat\F_{\nu i}^a\,,
\end{equation}
it is  apparent that in the condensed vacuum, a linear combination of $\F_{\mu i}^{a}+\widehat{\F}_{\mu i}^{a}$ enables us to identify the  BRST Goldstone degree of freedom. However,  the quantum version of the Ward identities \eqref{nodig} immediately lead to
\begin{equation}\label{brstino}
    \frac{\delta^2 \Gamma}{\delta \oF_{\mu i,x}^a \F_{\nu j,y}^b}=\frac{\delta^2 \Gamma}{\delta \widehat{\oF}_{\mu i,x}^a \widehat{\F}_{\nu j,y}^b}=-\delta_{\mu\nu, i j}^{ab}\p^2 \delta(x-y)\,,
\end{equation}
telling us that the Goldstone mode is exactly a free particle. As such, it can be trivially decoupled  from any consistent definition of a possible physical subspace.

Let us also point out that, relying on the observations that the spontaneous symmetry braking of the BRST symmetry can be encoded into the ($\h, \widehat{\h})$-EOMs, which have the meaning of  Ward identities for algebraic degrees of freedom, and that  the ensuing Goldstone degree of freedom itself is also completely characterized by a Ward identity, we can speak about an algebraic breaking of the BRST symmetry.

\section{Invariance under time evolution of the BRST operator cohomology}
\subsection{Preliminaries}
The aim of this section is to exploit some consequences of having at our disposal an exact BRST invariance which is spontaneously broken. In particular, the nilpotency of the BRST operator, eqs.\eqref{BRS},\eqref{BRS1}, opens the door to make use of cohomology tools in order to identify the BRST invariant local composite operators. From the well established results on the cohomology of Yang-Mills theories, see \cite{Barnich:2000zw,Piguet:1995er}, it follows that the cohomology classes in the space of zero ghost number are given by local colorless gauge invariant operators built up with the field strength $F^a_{\mu\nu}$ and its covariant derivative $D^{ab}_\mu$. Up to this point things look similar to the usual unbroken case. It is worth to emphasize there is, however, a deep difference between the unbroken and the broken case. The BRST invariant operators are always defined modulo exact BRST terms which, in the unbroken case, are physically irrelevant. This is not the case in the present situation. The exact BRST pieces might now give nontrivial contributions to the correlation functions, due to the non-invariance of the vacuum. A detailed study of how these exact terms contribute to a given correlation function is beyond the scope of the present letter. Though, as a first result, we shall give here a proof of the invariance under time evolution of the cohomology of the BRST operator in the spontaneously broken case. Denoting by $\{ \mathcal{O} \}$ the set of local BRST invariant operators, including the exact terms, it turns out that  $\{ \mathcal{O} \}$ is closed under time evolution, namely the evolution in time of $\{ \mathcal{O} \}$ will not bring us outside of $\{ \mathcal{O} \}$.

In order to give a proof of this statement, we shall make use of a toy model, already introduced in \cite{Baulieu:2009ha}, which reproduces the main features of the Gribov-Zwanziger theory without leading to notational clutter with indices etc. It is specified by the non-local Euclidean action
\begin{equation}\label{model0}
    S=\int \d^4x\; \frac{1}{2}\psi\left(-\p^2-\frac{2\theta^4}{\p^2}\right)\psi  \;,
\end{equation}
which is easily seen to give rise to a Gribov type propagator
\begin{equation}
\langle \psi(k)  \psi(-k) \rangle = \frac{k^2}{k^4+ 2\theta^4} \;.
\end{equation}
Despite its non-locality, the action  \eqref{model0}  can be cast in local form, which reads
\begin{equation}\label{model0b}
    S=\int \d^4x \left(\frac{1}{2}\psi(-\p^2)\psi+\overline\varphi(-\p^2)\varphi+\overline\omega (\p^2)\omega + \theta^2\psi(\varphi-\overline\varphi)\right)
\end{equation}
whereby we introduced 2 commuting $(\varphi, \overline\varphi)$ and 2 anti-commuting scalar fields $(\omega, \overline\omega)$, with BRST transformation
\begin{equation}\label{model2}
   s\psi=0\,,\qquad s\varphi=\omega\,,\qquad s\omega=0\,,\qquad s\overline\omega=\overline\varphi\,,\qquad s\overline\varphi=0\,.
\end{equation}
We notice that, as in the case of the Gribov-Zwanziger action   \eqref{GZaction},  the BRST symmetry is softly broken by the mass mixing term. Nevertheless, we can repeat the same procedure employed to construct the action  \eqref{GZaction2}, and reformulate expression \eqref{model0b} in a manifest BRST invariant fashion. As done before, we introduce an additional BRST quartet $(\F,\overline \F, \G,\overline \G)$, and a singlet $\h$, and consider
\begin{eqnarray}\label{model3}
    S&=&\int \d^4x \left(\frac{1}{2}\psi(-\p^2)\psi+\overline\varphi(-\p^2)\varphi+\overline\omega (\p^2)\omega + \theta^2\psi(\varphi-\overline\varphi)\right.\nonumber\\
    &&\left.+\overline \G(-\p^2)\G+\overline \F (\p^2)\F+\overline \G(\varphi-\overline\varphi)-\overline \F \psi\omega + \h(\overline \G-\theta^2)\vphantom{\frac 1 2}\right)\,,
\end{eqnarray}
with
\begin{equation}\label{model3b}
   s\G=\F\,,\qquad s\F=0\,,\qquad s\overline \F=\overline \G\,,\qquad s\overline \G=0\,,\qquad s\h=0\,.
\end{equation}
It is easily checked that $sS=0$, while the $\h$-EOM will set $\overline \G=\theta^2$, and by integrating out the $(\overline \F,\F)$ fields, equivalence at the quantum level with the original local action \eqref{model0b} is again established. Here, the $\F$-field plays the role of the Goldstone mode for the BRST breaking, which is linked to the exact quantum EOM,  $\frac{\delta \Gamma}{\delta \h}=0$. The BRST current reads
\begin{equation}\label{cur1}
    j_\mu = -\left(\p_\mu \overline \varphi\right)\omega + (\p_\mu\omega)\overline\varphi + (\p_\mu F)\overline G - (\p_\mu\overline G)F
\end{equation}
Let us mention that the action \eqref{model3} seems to harbor a local ``gauge'' symmetry, generated by
\begin{equation}\label{pi1}
    \delta_\epsilon \h(x) = \p^2 \epsilon(x)\,,\qquad \delta_\epsilon \G(x) = \epsilon(x)
\end{equation}
with $\epsilon(x)$ an infinitesimal function of $x$. This is actually corresponding to the following local Ward identity:
\begin{equation}\label{1}
    \p^2 \frac{\delta S}{\delta \h}+\frac{\delta S}{\delta \G}=0\,.
\end{equation}
It is also there in the reformulated version of the GZ action, one just needs to combine the identities \eqref{ward1}-\eqref{ward2}. In order to avoid to deal with first class constraints ---which typically correspond to a gauge freedom \cite{ht,Bekaert:2009zz}--- we shall canonically quantize the theory \eqref{model3} without the explicit use of the $\h$-field in the Hamiltonian formalism, and we shall  impose $\oG=\theta^2$ as the primary constraint to start from.

\subsection{A few words on Euclidean time evolution}
Strictly speaking, we are not allowed to Wick-rotate a Gribov-like action \eqref{model3} from Euclidean to Minkowski space. We need to work directly in Euclidean space. With the normal connection $x_0=-ix_0'$, where the prime refers to Euclidean space, we find that the Euclidean Heisenberg equation, which governs the time evolution of any operator $\mathcal{O}$, is given by
\begin{equation}\label{time1}
    \frac{\p}{\p x_0'} \mathcal{O} = -[\mathcal{O},\mathrm{H}]\,,
\end{equation}
with solution
\begin{equation}\label{time2}
   \mathcal{O}(x_0',\mathbf{x})= e^{x_0'\mathrm{H}}\mathcal{O}(\mathbf{x})e^{-x_0'\mathrm{H}}\,.
\end{equation}
Let us also have a look at the definition of the Euclidean Hamiltonian. We remind that the Euclidean action is given by
\begin{equation}\label{time3}
    \mathcal{L}'=-\mathcal{L}(\ldots_0=-i\ldots_0')
\end{equation}
where each time component needs to be transformed from Minkowski to Euclidean signature in the designated way. In Minkowski space-time, one introduces the conjugate momentum of a field\footnote{We consider a scalar field for simplicity here. In case of e.g.~a vector field, one needs to keep in mind to also transform the 0-component of the field itself in the appropriate way.} $\phi$ as
\begin{equation}\label{time4}
    \Pi  =  \frac{\delta \mathcal L}{\delta \dot{\phi}}= i~\frac{\delta \mathcal{L}'}{\delta\dot\phi'}=i~\Pi'   \;, \qquad
    \dot\phi  =  \frac{\p}{\p x_0}\phi\;,  \qquad
     \dot\phi'  =  \frac{\p}{\p x_0'}\phi \;,
\end{equation}
where we introduced the Euclidean conjugate momenta $\Pi'$ in a natural way from the Euclidean Lagrangian. For the Hamiltonian, we then find
\begin{equation}\label{time5}
    \mathrm{H}=  \int d \mathbf{x} \; \left( \dot\phi\Pi -\mathcal{L} \right) =  \int d \mathbf{x} \; \left( -\dot\phi'\Pi'+\mathcal{L}' \right) \;.
\end{equation}
At the level of the quantization, the usual equal-time-commutator reads
\begin{equation}\label{time6}
    \left[\phi(\mathbf{x},x_0),\Pi(\mathbf{y},x_0)\right]=i\delta(\mathbf{x}-\mathbf{y}) \;.
\end{equation}
Thus at the Euclidean level, we shall set
\begin{equation}\label{time6}
    \left[\phi'(\mathbf{x},x_0'),\Pi'(\mathbf{y},x_0')\right]=\delta(\mathbf{x}-\mathbf{y}) \;.
\end{equation}
\subsection{A few words on the definition of the charge}
Sometimes one reads that in case of spontaneous symmetry breaking, associated to a conserved current $J_\mu$, the following definition of the charge\,,
\begin{equation}\label{charge1}
    \Q=\int \d^3x \;J_0\,,
\end{equation}
fails. This is true, in the sense that it is not possible to define a global version of the charge, since the behavior of the massless modes at infinity jeopardizes the convergence of the integral. However, it should still be possible to consider the local variation of any quantity w.r.t.~the transformation behind the conserved current\footnote{Or even more generally, to define the variation of e.g.~the Hamiltonian under any transformation, symmetry-related or not.}. As explained in Ch.~17 of \cite{Strocchi:2005yk}, one defines
\begin{equation}\label{charge2}
    \Q^R=\int \d^4 x\; f^R(\mathbf{x}) \alpha(x_0) J_0(\mathbf{x},x_0)
\end{equation}
where the local action of the transformation acting on the operator $\mathcal{O}$ is obtained from
\begin{equation}\label{charge2b}
    \delta\mathcal{O}=\lim_{R\to\infty}[\Q^R,\mathcal{O}]\,.
\end{equation}
In the above, one sets
\begin{equation}\label{charge3}
    f^R(\mathbf{x})=f(|\mathbf{x}|/R)\,,\qquad f(y)=1\quad\text{if }|y|\leq 1\,,\qquad f(y)=0\quad\text{if }y> 1
\end{equation}
and for $\alpha(x_0)$ one may take any smooth approximation of the $\delta$-function, keeping $\int \d x_0 \alpha(x_0)=1$. The presence of  $\alpha(x_0)$ is a technical issue. It amounts to a smearing in time, a feature needed for a mathematically well defined object in the case of relativistic QFT, in the strict Wightman-axiomatic approach to QFT. We shall not be concerned with this smearing in this note\footnote{One shows that the limit $R\to\infty$ exists, and that it is independent of the choice of the smearing function $\alpha(x_0)$ \cite{Strocchi:2005yk}.}. We shall  tacitly assume its presence  and therefore simply write $\int \d^3x f^R(\mathbf{x})$. The most important part of the foregoing construction is the presence of a cut-off at large $|\mathbf{x}|$ of the integrand, avoiding the potential problem of long-wavelength divergences caused by the massless modes.

Thus, having introduced $\Q^R$ and the transformation law \eqref{charge2b}, we are equipped to define an operator subspace $\mathfrak{F}$ subject to
\begin{equation}\label{sub1}
    \mathcal{O}\in\mathfrak{F}\Leftrightarrow [\Q,\mathcal{O}]=\lim_{R\to\infty}[\Q^R,\mathcal{O}]=0\,.
\end{equation}
This subspace is then invariant under time evolution.  Indeed, we easily find, using \eqref{time2} and dropping from now on the prime-notation for Euclidean quantities,
\begin{equation}\label{sub2}
    \lim_{R\to\infty}[\Q^R,e^{x_0\mathrm{H}}\mathcal{O}e^{-x_0\mathrm{H}}]=e^{x_0\mathrm{H}} \left(\lim_{R\to\infty}[\Q^R,\mathcal{O}]\right)e^{-x_0\mathrm{H}}=0\,,
\end{equation}
as $\mathcal{O}\in\mathfrak{F}$. If $\Q^R$ happens to be a nilpotent operator, the operator cohomology is invariant under time evolution. Notice that nowhere in this argument we used that $\lim_{R\to\infty}\Q^R$ corresponds to a broken or unbroken symmetry generator, in the sense that we did not use the fact that there exists (or not) an operator $\mathcal{A}$ with $\Braket{\lim_{R\to\infty} [\Q^R,\mathcal{A}]}\neq 0$.

It is only at the level of taking expectation values of products of operators belonging to $\mathfrak{F}$ that the effects of the broken symmetry will emerge, as the expectation values need to be taken w.r.t~the correct vacuum. For instance,  as stated previously, in the case of  the spontaneous symmetry breaking of the BRST symmetry, the choice of the nontrivial vacuum will have the consequence that in expressions of the form
\begin{equation}\label{sub3}
    \Braket{\left[\mathcal{A}_1(x)+s\mathcal{A}_2(x)\right]\left[\mathcal{A}_3(y)+s\mathcal{A}_4(y)\right] }\,,
\end{equation}
the  $s$-exact pieces will give a nonvanishing contribution to the correlation function.

\subsection{Back to the quantization of the toy model}
Using the tools of the previous section, we first derive the conjugate momenta,
\begin{eqnarray}\label{PB0}
  \Pi_\psi &=& \p_0\psi\,,\qquad \Pi_\varphi=\p_0\overline\varphi\,,\qquad \Pi_{\overline\varphi}=\p_0\varphi\,,\qquad \Pi_\omega=-\p_0\overline\omega\,,\qquad   \Pi_{\overline\omega}=\p_0\omega\,,\nonumber\\
   \Pi_\F&=&-\p_0\overline \F\,,\qquad  \Pi_{\overline \F}=\p_0 \F\,,\qquad  \Pi_\G=\p_0 \G\,,\qquad  \Pi_{\overline \G}=\p_0 \G\,.
\end{eqnarray}
Next, we impose $\oG-\theta^2=0$ as a so-called primary constraint \cite{ht,Bekaert:2009zz}. Consistency of this constraint with time evolution\footnote{This amounts to having a weakly vanishing ($\approx 0$) Poisson bracket of the constraint(s) with the Hamiltonian, where weakly means modulo the previous constraints.} will give rise to higher order ``secondary'' constraints. Therefore, we need the Hamiltonian, which is found to be
\begin{eqnarray}\label{PB1}
    \mathrm{H}&=&\int \d^3x \left[-\frac{1}{2}\Pi_\psi^2-\Pi_\phi\Pi_{\overline\phi}-\Pi_\omega\Pi_{\overline\omega}-\Pi_\F\Pi_{\overline \F}-\Pi_\G\Pi_{\overline \G}+\frac{1}{2}(\p_i\psi)^2+\p_i\varphi\p_i\overline\varphi+\p_i\omega\p_i\overline\omega\right.\\&&\left. \;\;\;\;+\p_i \G\p_i\overline \G +\p_i \F \p_i\overline \F + \overline \G\psi(\varphi-\overline\varphi)-\oF \psi \omega \right]
\end{eqnarray}
and the Poisson bracket, which is in general given by
\begin{equation}\label{PB2}
    \left\{A(\mathbf{x},t),B(\mathbf{y},t)\right\}= \sum_{\phi}\int \d^3 z \left[\frac{\delta A}{\delta \phi(\mathbf{z},t)}\frac{\delta B}{\delta \Pi_\phi(\mathbf{z},t)}-\frac{\delta A}{\delta \Pi_\phi(\mathbf{z},t)}\frac{\delta B}{\delta \phi(\mathbf{z},t)}\right]\,,
\end{equation}
with $t=x_0$ identified as the Euclidean time. As we shall see, there will be no first-class constraints\footnote{First-class constraints are those with vanishing Poisson bracket with all other constraints. Second-class constraints are then the remaining ones.}.

Let us now construct the secondary constraints. From $\oG-\theta^2=0$, we get $\Pi_\G=0$ via $\left\{\oG,\mathrm{H}\right\}\approx 0$. Also, when $\oG=\theta^2$, we have simultaneously $\Pi_{\oG}\approx 0$ which, in turn, leads to $\p_i^2 \G=0$ via $\left\{\Pi_{\oG},\mathrm{H}\right\}\approx 0$. Here, we can stop since $\left\{ \p_i^2 \G,\mathrm{H}\right\}\propto \p^2\Pi_{\oG}$ and the latter is indeed weakly zero due to a constraint  already found. Likewise, $\left\{\Pi_\G,\mathrm{H}\right\}\propto -\p_i^2\oG\approx 0$, not providing any new information.

The constraints $\oG-\theta^2$, $\Pi_{\overline{\G}}$ and $\p_i^2 \G$, $\Pi_\G$ are pairwise second class, as a quick computation learns that $\left\{\oG(\mathbf{x},t)-\theta^2,\Pi_{\oG}(\mathbf{y},t)\right\}=\delta(\mathbf{x}-\mathbf{y})$ and $\left\{\p_i^2 \G(\mathbf{x},t),\Pi_{\G}(\mathbf{y},t)\right\}=\p_i^2\delta(\mathbf{x}-\mathbf{y})$.  Any other bracket of constraints is vanishing.

In order to quantize the theory, we first need to replace the Poisson bracket with its more general Dirac counterpart \cite{ht,Bekaert:2009zz}. In general, the Dirac bracket is defined as\footnote{For notational simplicity, we dropped the (equal) time $t$ everywhere.}
\begin{equation}\label{PB5}
    \left\{A(\mathbf{x}),B(\mathbf{y})\right\}^*=     \left\{A(\mathbf{x}),B(\mathbf{y})\right\} - \sum_{\alpha,\beta}\int \d^3 z\int \d^3z'\left\{A(\mathbf{x}),\Phi_\alpha(\mathbf{z})\right\} \mathcal{M}_{\alpha\beta}^{-1}(\mathbf{z},\mathbf{z}')\left\{\Phi_\beta(\mathbf{z}'),B(\mathbf{y})\right\}\,.
\end{equation}
Here, $\Phi_\alpha$ are the second class constraints, with the anti-symmetric invertible matrix $\mathcal{M}_{\alpha\beta}=\left\{\Phi_\alpha,\Phi_\beta\right\}$. In the above case, using the entries $\oG$, $\Pi_{\oG}$, $\p_i^2 \G$, $\Pi_\G$, this matrix and its inverse are found to be
\begin{equation}\label{PB6}
    \mathcal{M}=\left(
                                  \begin{array}{cccc}
                                    0 & \delta(\mathbf{z}-\mathbf{z}') & 0 & 0 \\
                                    -\delta(\mathbf{z}-\mathbf{z}') & 0 & 0 & 0 \\
                                    0 & 0 & 0 & \p_i^2 \delta(\mathbf{z}-\mathbf{z}') \\
                                    0 & 0 & -\p_i^2 \delta(\mathbf{z}-\mathbf{z}') & 0 \\
                                  \end{array}
                                \right)\,,\qquad \mathcal{M}^{-1}=\left(
                                  \begin{array}{cccc}
                                    0 & -\delta(\mathbf{z}-\mathbf{z}') & 0 & 0 \\
                                    \delta(\mathbf{z}-\mathbf{z}') & 0 & 0 & 0 \\
                                    0 & 0 & 0 & \frac{1}{4\pi}\frac{1}{|\mathbf{z}-\mathbf{z}'|} \\
                                    0 & 0 & -\frac{1}{4\pi}\frac{1}{|\mathbf{z}-\mathbf{z}'|} & 0 \\
                                  \end{array}
                                \right)\,.
\end{equation}
This leads to the usual brackets for all fields, with the exception of
\begin{equation}\label{PB7}
    \left\{ \oG(\mathbf{x}),\Pi_{\overline{G}}(\mathbf{y})\right\}^*=0\,,\qquad \left\{ \G(\mathbf{x}),\Pi_\G(\mathbf{y})\right\}^*=0
\end{equation}
as follows after a little computation based on \eqref{PB5},\eqref{PB6}. Here, we clearly see the purpose of the Dirac bracket: it becomes zero whenever it contains a 2nd class constraint. In order to complete the quantization of the model, we must replace the Dirac brackets with (anti)commutators, according to the prescription \eqref{time6}. Since the Dirac bracket of a constraint automatically vanishes, no inconsistencies will be retrieved. In fact, one has the choice to first use the constraints and then compute the bracket, or vice versa. Concretely, we have the following (anti-)commutators
\begin{eqnarray}\label{DB1}
    \left[\phi(\mathbf{x},t),\Pi_\phi(\mathbf{y},t)\right]_{\pm}&=&\delta(\mathbf{x}-\mathbf{y})\,,\qquad \text{for } \phi\in\left\{\psi,\varphi,\overline\varphi,\omega,\overline\omega, \F, \overline \F\right\}\,,\nonumber\\
    \left[\G(\mathbf{x},t),\Pi_\G(\mathbf{y},t)\right]&=&\left[\oG(\mathbf{x},t),\Pi_{\oG}(\mathbf{y},t)\right]=0\,.
\end{eqnarray}
As a last step, we need to establish that the Hamiltonian still commutes with the BRST charge, for which we rely on
\begin{equation}\label{PB15}
    \Q^R= \int \d^3 x f^R(\mathbf{x}) \left[-\Pi_\varphi \omega + \Pi_{\overline\omega}\overline\varphi+\Pi_{\overline{\F}} \overline{\G} - \Pi_{\G}\F\right]\,.
\end{equation}
After some algebra, one finds
\begin{eqnarray}\label{PB16}
    [\Q^R,\mathrm{H}]&=& \int \d^3x f^R(\mathbf{x}) \left(\Pi_{\oF}(\mathbf{x},t)\Pi_{\G}(\mathbf{x},t)+\p_i \F(\mathbf{x},t) \p_i \oG(\mathbf{x},t)\right]\nonumber\\
    &\approx&0\,,
\end{eqnarray}
upon using the constraints. The same result is obtained if the constraints are used before evaluating the bracket.

\section{Conclusion}
We believe the essence of the novel formulation of the Gribov-Zwanziger action presented here is that:  (1) we can introduce the cohomology of the BRST operator which enables us to renormalize quantum extensions of the classically gauge invariant operators, (2) the spontaneous symmetry breaking of the BRST symmetry can be completely kept under control in an algebraic way, with a decoupled Goldstone mode, and (3) the operator cohomology forms a closed subspace under time-evolution, despite that the BRST symmetry is algebraically broken in the true vacuum. The ``only'' noticeable consequence of this breaking will be the nonvanishing of correlation functions of BRST exact quantities.  One point we, willingly, did not touch upon in this letter is what happens in the case of spontaneously broken BRST symmetry with the gauge parameter independence of correlation functions of quantities that belong to the BRST cohomology. Here, we cannot provide an answer, as the Gribov-Zwanziger construction is \emph{strictly} limited to the Landau gauge\footnote{There is also a version in the Coulomb gauge, but the latter gauge makes e.g.~the renormalization analysis and loop computations hard, not to say impossible, because of its noncovariant nature.}. In particular, the defining properties of the Landau gauge are heavily used in the derivation of the partition function. For example, a key role is played by the Hermitian character of the Faddeev-Popov operator, something which holds only in the case of the Landau gauge. As such, in the absence of a Gribov-Zwanziger construction for e.g.~a general linear covariant gauge, it is meaningless  trying to establish the gauge parameter independence of certain quantities. One could easily imagine the Gribov parameter to become gauge parameter dependent in such a setting, if the underlying action would be known to begin with. Nonetheless, if someday somebody would manage to construct a sensible Gribov-Zwanziger action with gauge parameters, the exact Slavnov-Taylor identity established here could be used to derive a kind of Nielsen identities \cite{DelCima:1999gg} to control the gauge parameter dependence.

At this point, we have reached the same conclusion as anybody can reach using the QCD action in a particular gauge which would have perfect nilpotent BRST invariance: we have identified a well-defined subspace of renormalizable quantum extensions of the classically gauge invariant operators. This is the first step in the fundamental issue of establishing \emph{unitarity} of the confined degrees of freedom. The second, highly nontrivial step, is to check the positivity in this subspace. This has in se nothing to do with having the BRST at one's disposal or not. From parts of the existing literature, sometimes unattentive readers might get the impression that having a nilpotent BRST charge is sufficient to ensure unitary of the underlying field theory, as long as the BRST is a symmetry. This is actually never true, even  at the perturbative level. Indeed, one always needs to check that the degrees of freedom singled out by the BRST cohomology do have positive norm. This is, by the way, clearly mentioned in the classic work \cite{Kugo:1979gm}, see the discussion below eq.~(3.6) on p.~25. An explicit counterexample of a gauge theory with a nilpotent BRST invariance which is not unitary in terms of its (asymptotic) elementary degrees of freedom can be found in e.g.~\cite{Dudal:2007ch}.  Things get even more complicated in the context of a confining gauge theory: what if one cannot define the asymptotic states of the elementary excitations when the latter are confined? According to our understanding, one then needs to access the spectral properties of the two point correlation functions of composite operators, constructed from their confined constituents. It is now  clear that the BRST symmetry itself does not provide any answer to such inherently nonperturbative spectral question. In fact, it appears that very little is concretely known about the spectral properties of glueballs or hadrons in general. For example, lattice QCD offers a powerful way to access gauge theories nonperturbatively, but from a Monte Carlo simulation in Euclidean space time at momentum $p^2>0$, it is not straightforward to extract anything on the behaviour of e.g.~two-points functions at complex momenta. Also, most functional approaches to the bound state problem, viz.~a combination of Dyson-Schwinger and Bethe-Salpeter equations, are usually not focused on obtaining information in the whole complex momentum plane. Anyhow, given the nontrivial analytical structure of the input Green functions of gluons, ghosts and quarks, it is to be expected that it will \emph{not} be a trivial matter  to assure the correct spectral properties of their bound states. For example, propagators with complex conjugate poles lead to cuts in the complex plane when these poles are  suitably combined, see e.g.~\cite{Baulieu:2009ha} for an exploratory study in a Gribov-like toy model.

In order to avoid dealing directly with the spectral representation of the correlation functions of composite operators, an alternative approach would be to construct, directly from the QCD action, the \emph{effective Hamiltonian} for e.g.~the glueball degrees of freedom, and to carry out a canonical quantization with the latter Hamiltonian. Unfortunately, as far as we know, this is far beyond the current capabilities. In general, we are talking here about particles with higher spin, notoriously hard to handle in a self-consistent relativistic quantum field theory way. The next best thing would be a non-relativistic Hamiltonian description. But also here, who knows how to construct even this ab initio from the QCD action?

To end, we believe a lot of work needs to be carried out before a full satisfactory understanding of the complicated issue of unitarity in a confining gauge theory will become available. With this letter, we hope to have provided insight into at least one face of this problem: even with a spontaneous breaking of the BRST as is the case in the nonperturbative Gribov-Zwanziger Landau gauge quantization, a consistent definition of a time invariant operator subspace can still be given, very much alike as in the case of unbroken BRST. As a consequence, the unitarity question in both unbroken and broken case boils down to the same difficulty: to establish whether the composite operator Green functions do display the correct analytical properties to represent true physical degrees of freedom.

\end{document}